\documentclass{bioinfo}
\usepackage{amssymb, amsmath}
\usepackage{url}
\copyrightyear{2015} \pubyear{2015}

\access{Advance Access Publication Date: Day Month Year}
\appnotes{Applications Note}

\begin{document}
\firstpage{1}

\subtitle{Genome analysis}

\title[H(O)TA]{H(O)TA: estimation of DNA methylation and hydroxylation levels and efficiencies from time course data}
\author[Kyriakopoulos, Giehr and Wolf]{Charalampos Kyriakopoulos\,$^{\text{\sfb 1,}*}$, Pascal Giehr\,$^{\text{\sf 2}}$ and Verena Wolf\,$^{\text{\sfb 1,}*}$}
\address{$^{\text{\sf 1}}$Computer Science Department, Saarland University, Saarbruecken, 66123, Germany and \\ $^{\text{\sf 2}}$Department of Biological Sciences, Saarland University, Saarabruecken, 66123, Germany.}

\corresp{$^\ast$To whom correspondence should be addressed.}

\history{Received on XXXXX; revised on XXXXX; accepted on XXXXX}

\editor{Associate Editor: XXXXXXX}

\abstract{\textbf{Motivation:} Methylation and hydroxylation of cytosines to form 
5-methylcytosine (5mC) and
5-hydroxymethylcytosine (5hmC) belong to the most important
epigenetic  modifications  
 and their vital role in the  regulation of gene expression has been widely recognized.
  Recent experimental techniques  allow to infer
 methylation and hydroxylation levels  at CpG dinucleotides
 but require a sophisticated statistical analysis to 
 achieve accurate estimates.\\
\textbf{Results:} We present H(O)TA, a software tool based on a stochastic modeling approach, which 
  simultaneously  analyzes time course data from hairpin bisulfite sequencing
and  hairpin oxidative bisulfite sequencing.  \\
\textbf{Availability and additional information:} \url{https://mosi.uni-saarland.de/HOTA} \\
\textbf{Contact:}  \href{charalampos.kyriakopoulos@uni-saarland.de}{charalampos.kyriakopoulos@uni-saarland.de} or \href{verena.wolf@uni-saarland.de}{verena.wolf@uni-saarland.de}\\
}

\maketitle

\section{Introduction}
DNA methylation  refers to the  transfer of a methyl group to the C-5 position of cytosine (C) to produce 5-methylcytosine (5mC).  In mammals it is predominantly found in the  symmetrical CpG context  and, as a major epigenetic modification,
  it plays an essential role in the regulation of gene expression.
  Moreover,  DNA methylation contributes to a wide range of cellular processes
  such as development, X-inactivation, and imprinting (\cite{bourc2004meiotic}) 
 and aberrant methylation patterns have been linked to several human diseases including cancer (\cite{herman1999hypermethylation}). 
 The oxidized form, 5-hydroxymethylcytosine (5hmC),
 has recently gained attention as it  is not only involved in gene regulation but also seems to  play a major role in active and passive DNA demethylation.  It is hypothesized that CpGs can traverse an iterative cycle of methylation and demethylation through oxidation and   base excision repair (\cite{zhang2012thymine}). 

\begin{figure}[t]
\centerline{\includegraphics[width=0.4\textwidth]{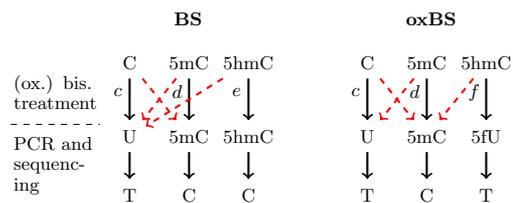}}
\caption{Conversion scheme for BS-seq and oxBS-seq. The first (last) line determines the hidden (observable) states. The conversion rates $c,$ $d$, $e$, and $f$
are taken into account by adjusting the emission probabilities
of the models. }\label{fig:converror}
\vspace*{-10pt}
\end{figure}

DNA methylation is commonly measured by using bisulfite genomic sequencing (BS-seq) during which C is converted to uracil (\cite{frommer1992genomic})
while  both 5mC and 5hmC are read as Cs and  can therefore not be discriminated (see Fig.~\ref{fig:converror}).  
As opposed to this, oxidative bisulfite sequencing 
(oxBS-seq) converts   5hmC to 5-formylcytosine (5fC) and conversion of the newly formed 5fC to uracil allows to   discriminate between 5hmC and 5mC~but not between   C and 5hmC (\cite{booth2012quantitative}).
Hence  5hmC levels must be inferred by a simultaneous
estimation based on both BS-seq and  oxBS-seq data.
Standard BS-seq or oxBS-seq can only capture the modification state of one individual DNA strand at a time. To overcome this limitation hairpin BS-seq has been developed,
which allows to determine the state of both cytosines of a CpG dyad.
Thus,  nine different possible
states (pairs of the three  possible states C, 5mC and 5hmC)
can be implicitly measured (\cite{laird2004hairpin}).
Moreover, while most cell types display relatively stable DNA methylation patterns,
  the dynamically changing gene expression program during mammalian development is accompanied by an alteration of methylation patterns. 
Given measurements at different times,  (time-dependent) methylation
  efficiencies can be inferred
     and   provide useful information
about the mechanisms that control the developmental program (\cite{arand2012vivo}). 
 
 Here, we present \underline{H}airpin (\underline{O}xidative) bisulfite sequencing \underline{T}ime course \underline{A}nalyzer (H(O)TA) -   a tool that  accurately infers (hydroxy-)methylation levels and determines the efficiencies of the involved enzymes at a certain DNA locus.
 The procedure for estimating levels and efficiencies is based on  the construction of two coupled Hidden Markov Models (HMMs) and gets as input time course measurements from hairpin  BS-seq and  oxBS-seq,
 respectively. Using the ML approach proposed in \cite{giehr2016}, 
   the evolution of the HMMs is determined by time-dependent methylation and (hydroxy-)\break methylation efficiencies and  takes into account all relevant conversion errors (see dashed arrows
 in Fig.~\ref{fig:converror}).


\vspace*{-14pt}

\section{H(O)TA Software}

\begin{figure}[t]
       	\centerline{\includegraphics[width=0.5\textwidth]{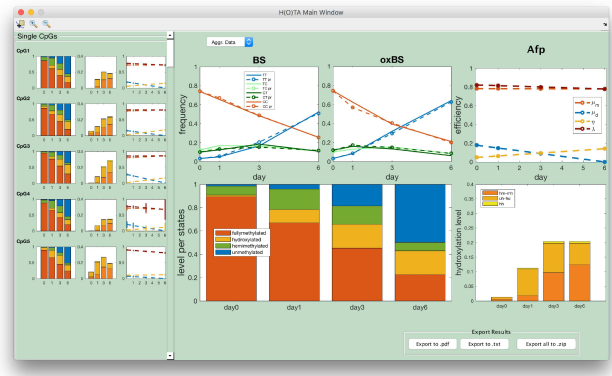}}
       	\caption{The main window of the graphical user interface of H(O)TA.\label{fig:main} }
       	\vspace*{-10pt}
       \end{figure}

 H(O)TA has been developed in MATLAB 
 and its execution requires the installation of
 the free Matlab runtime environment (MRE). The
 tool and the MRE can be downloaded as a single   installation file  available for Linux, MacOS, and Windows operating systems.
As opposed to   methods for single time point data,
    H(O)TA performs an analysis that considers the transient probability
     distribution over the set $\{u,m,h\}^2$ of nine hidden states  of the two cytosines of a CpG dyad, where $u, m$ and $h$ describe    C,   5mC, and    5hmC, respectively.  Thus, besides the   states $uu$  and $mm$, which correspond to the blue and red bars in the   bar plots of the hidden states' probabilities
      in Fig.~\ref{fig:main}, the model considers 
     hemimethylated sites (states    $um$, $mu$, green bars) as well as fully hydroxylated sites
    (state $hh$)  
    and hemihydroxylated sites (states
      $uh$, $hu$) and combinations of 5mC and 5hmC (states        $mh$, $hm$), whose levels are given by
       orange bars and refined in detailed plots on the right of each bar plot in Fig.~\ref{fig:main}.
                The observable states reflect the possible outcomes of hairpin BS-seq
   and hairpin oxBS-seq, respectively, that is, $\{T,C\}^2$ (cf. 
   last line in Fig.~\ref{fig:converror} and upper middle line plots).
   Users can provide BS-seq and oxBS-seq time course data.
 For each observation time point, estimates of the  methylation and hydroxylation 
     levels are computed, as well as,  linear functions  
    for the methylation (maintenance or de novo) and hydroxylation
    efficiencies, i.e.,  the probability
    of a methylation or a hydroxylation event between two
    cell divisions. In addition, an estimation is provided for the  probability that  no maintenance is performed when the current state is $mh$ or $hm$, which hints on  the existence of a  passive demethylation mechanism induced by hydroxylation.
   The user must specify
     the number of cell divisions between two observation time points and provide   conversion errors of
     BS-seq and oxBS-seq (see dashed arrows in Fig.~\ref{fig:converror}). 
   For all the estimated parameters confidence intervals
   are computed and a  statistical test is carried out in order to verify certain hypotheses about the   efficiencies. 
   For a derivation of   the likelihood   and details about the  optimization as well as the statistical validation of the results, we refer to (\cite{giehr2016}).


The graphical user interface of H(O)TA consists of two windows: a dialogue window for loading the input files of a DNA locus and running the analysis and the main window (Fig.~\ref{fig:main}) for visualizing the output.
The tool can automatically aggregate data of   different CpGs of a locus and compute average (hydroxy-)methylation levels as well as average efficiencies.
In addition, the analysis  can be performed for each CpG individually.

Users can provide three input .txt files. The first file contains BS-seq data, the second one  oxBS-seq data and the third file   contains the conversion errors of the two experiments as well as a string that describes  how many cell divisions take place between two  observation time points. Only the  file with the BS-seq data is mandatory and
the other two are optional. If only BS-seq data is given,
then the tool will predict only the methylation levels of the region (merged with the unknown hydroxylation levels).
For a detailed documentation of the input files we refer to the tool webpage. 

The main window has two subpanels. The right panel shows the output of the analysis either for the aggregated data or for each of the previously chosen CpG sites. 
The upper left and middle plots show the fit between the data and the model prediction for the observable states TT, TC, CT, CC for BS and oxBS experiments. 
The upper right plot   presents the efficiencies of the enzymes responsible for maintenance methylation (dark red), de novo methylation
(blue) and hydroxylation (yellow) as well as the total methylation (light red) on hemimethylated CpGs (see the tool webpage). 
The lower left plot shows the (hydroxy-)methylation levels of the current region and the lower right plot shows the exact distribution of the different hydroxylation states.
 In the upper left corner of the right panel the user can choose between the plots of different CpG sites (or the aggregated data).
In the lower right corner there are several options for exporting the estimation results in a desirable format.
 In the left panel of the main window  the (hydroxy-)methylation levels and the efficiencies of all individual CpGs are plotted such that they can be compared with each other and with the chosen plots in the right panel.

%
%

\vspace*{-14pt}

%
%
%
%

\bibliographystyle{natbib}
\bibliography{document}

\end{document}